\documentclass[useAMS,usenatbib,twocolumn]{mnras}
\usepackage{graphicx}
\usepackage{dcolumn}   
\usepackage{bm}        
\usepackage{amssymb}   
\usepackage[dvipsnames]{xcolor}

\usepackage{bm}

\def\bq{\begin{equation}}
\def\eq{\end{equation}}
\def\bqy{\begin{eqnarray}}
\def\eqy{\end{eqnarray}}

\def\ga{\gamma}

\def\la{\lambda}

\def\p{\partial}

\def\rh{\rho}

\def\p{\partial}

\def\calf{\mathcal{F}}

\def\calj{\mathcal{J}}

\begin{document}

\title[MRI stresses in Hall MHD]{A heuristic model for MRI turbulent stresses in Hall MHD}
\author[M. Lingam and A. Bhattacharjee]{Manasvi Lingam$^{1}$\thanks{E-mail:
mlingam@princeton.edu} and Amitava Bhattacharjee$^{1,2}$\thanks{E-mail:
abhattac@pppl.gov}\\
$^{1}$Department of Astrophysical Sciences and Princeton Plasma Physics Laboratory, Princeton University, Princeton, NJ 08543, USA\\
$^{2}$Max Planck-Princeton Center for Plasma Physics, Princeton University, Princeton, NJ 08544, USA}

\pagerange{\pageref{firstpage}--\pageref{lastpage}} \pubyear{2016}

\maketitle

\label{firstpage}

\date{}

\begin{abstract}
Although the Shakura-Sunyaev $\alpha$ viscosity prescription has been highly successful in characterizing myriad astrophysical environments, it has proven to be partly inadequate in modelling turbulent stresses driven by the MRI. Hence, we adopt the approach employed by \citet{GIO03}, but in the context of Hall magnetohydrodynamics (MHD), to study MRI turbulence. We utilize the exact evolution equations for the stresses, and the non-linear terms are closed through the invocation of dimensional analysis and physical considerations. We demonstrate that the inclusion of the Hall term leads to non-trivial results, including the modification of the Reynolds and Maxwell stresses, as well as the (asymptotic) non-equipartition between the kinetic and magnetic energies; the latter issue is also addressed via the analysis of non-linear waves. The asymptotic ratio of the kinetic and magnetic energies is shown to be \emph{independent} of the choice of initial conditions, but it is governed by the \emph{Hall parameter}. We contrast our model with the Kazantsev prescription from small-scale dynamo theory, and the Hall term does not contribute in the latter approach, illustrating the limitations of the Kazantsev formalism. We indicate potential astrophysical applications of our model, including the solar wind where a lack of equipartition has been observed.
\end{abstract}

\begin{keywords}
(magnetohydrodynamics) MHD -- methods: analytical -- plasmas -- magnetic fields -- turbulence -- instabilities
\end{keywords}

\section{Introduction}
The magnetorotational instability (MRI) has proven to be a strikingly reliable and ubiquitous means of understanding the properties of a plethora of astrophysical systems. The MRI was first studied in the 1950s and 1960s \citep{V59,C60}, but its true potential in astrophysics was first realized by Balbus and Hawley in their pioneering work \citep{BH91}. The greatest success of the MRI has, arguably, been in the realm of modelling momentum transport in accretion discs \citep{HGB95,SHGB96,Gam96,Arm98,BH98,FSH00,FKR02,Bal03,SITS04,Bai11,BS13}. The MRI has subsequently been invoked in a number of settings, far too numerous to mention in detail, but we note that it has been successfully employed in dynamo theory \citep{TP92,SHGB96,VB97,Haw00,KHOTW04,BS05,Gress10,KK11,SB15}, launching of jets and winds \citep{MBA06,SI09,LFO13}, protoplanetary discs \citep{SMUN00,SW05,KreLin07,TSD07,KL10,OH11,BS13}, core-collapse supernovae and protoneutron stars \citep{AWML03,KSYS04,SLSS06,MBA06,Metal14}, zonal flows \citep{JYK09,FNKTH11,BS14}, etc. Experimental studies of the MRI are also widely prevalent in the scientific literature, see e.g. \citet{JGK01,GJ02,Setal04,Stef06,Stef07,SJB09,Set14}.

Given the success and versatility of the MRI, an immediate question that arises is the issue of transport coefficients, viz. the means of quantifying the enhanced momentum transport enabled by this instability. We note that the properties of these coefficients, as well as the operational limits, convergence, saturation and aspect ratio dependence (in the shearing box scenario) of the MRI-driven turbulence, remains an active and unresolved area of research \citep{FP07a,FPLH07,LL07,BMCRF08,BPV08,PG09,LLB09,FSt09,DSP10,LL10,SKH10,BS13,MFLJL15,SSH15}. The extent of MRI turbulence (the enhanced momentum transport) is captured by the use of the seminal $\alpha$ viscosity prescription, first proposed by Shakura and Sunyaev in \citet{SS73}. The $\alpha$ viscosity prescription had already been studied widely in the context of accretion discs before the advent of MRI \citep{NT73,SS76,Pac78,Pr81,LP87}, and has been subsequently utilized for the same purposes, but with the incorporation of MRI-related physics \citep{TP92,BPap99,FKR02,Bal03,KPL07}. Given the widespread usage of the $\alpha$ viscosity in characterizing MRI turbulence, a natural question that emerges is: how accurate is the former in fully encapsulating all the traits of the latter? 

It was argued in \citet{GIO03} that the $\alpha$ viscosity does not faithfully reproduce the complexities of the MRI turbulent stresses. By working in the domain of the shearing box, \citet{GIO03} proposed a partly heuristic model that captured effects such as the linear interactions of the stresses (Maxwell and Reynolds) with an underlying mean flow. Alternative models for the MRI turbulent stresses include the formulations of \citet{PCP06} and \citet{KY93,KY95}; the latter employed the Two Scale Direct Interaction Approximation \citep{YIIY01}. \citet{PCP07} showed that the large values of $\alpha$ measured from observations could be reproduced by numerical simulations of turbulent stresses only when specific conditions were fulfilled. Moreover, further work undertaken by \citet{PCP08,HB08,BN15} demonstrated that the standard shear viscosity paradigm was untenable, and emphasized the need for more sophisticated models of MRI turbulent stresses, such as the ones developed by \citet{GIO03} and \citet{PCP06}. The recent study by \cite{RLG16} has also pointed out the inherent ambiguities in the stress-pressure scaling that stem from the dual issues of numerical convergence and non-zero magnetic flux. In this paper, we shall attempt to extend the work of \citet{GIO03} to include a non-ideal magnetohydrodynamics (MHD) effect, by employing Hall MHD as our base model. In order to validate our decision, the effectiveness of the original model \citep{GIO03} must be established, and to make a convincing case for Hall MHD -- these factors are tackled below.

After the model for MRI turbulent stresses was proposed by \citet{GIO03}, it was subsequently employed quite successfully to study turbulent shear flows in \citet{GO05}, and stellar convection in \citet{MG07,GOMS10}. Moreover, the \citet{GIO03} model has also exhibited a reasonable degree of accuracy when compared against numerical simulations, as shown by the likes of \citet{LKKBL09,SBKM12}; see also \citet{KB08,SKKL09,SRKMB12} for associated discussions of the \citet{GIO03} model. However, it is also important to emphasize that models based on \citet{GIO03}, such as \citet{GOMS10}, are quite simple, and cannot reproduce all the features observed in direct numerical simulations, as pointed out in \citet{SKKR15}.

Now, let us consider the importance of Hall MHD in astrophysics. Hall MHD has been widely evoked in a diverse array of astrophysical contexts such as dynamos \citep{MGM02,MGM03a}, jets \citep{Bai14,LM15,Bai15}, protoplanetary discs \citep{W07,A11}, neutron stars \citep{GR92,CAZ04,GC14}, star formation \citep{W04,LKS11,KLS11}, and the turbulent properties of the solar wind \citep{GSRG96,KM04,GBuc07,ACVS08}. When it comes to the MRI, Hall MHD has been extensively investigated through a combination of analytic and numerical studies \citep{W99,BT01,SS02a,SS02b,ELFB11,Bai11,WS12,KL13,LKF14,Bai14,Bai15,SLKA15}. 

The pioneering work of \citet{SS02b} showed that the Hall parameter (ratio of the ion skin depth to the scale length) could either suppress or enhance the MRI turbulence, depending on the sign of the initial magnetic field. We find that our model is capable of supporting enhanced values of the saturated Maxwell stress for a positive Hall parameter, in agreement with the results obtained in \citet{SS02b}. Our work is broadly consistent with the notion that the Hall effect can play a key role in regenerating ``dead zones'' in protoplanetary discs, as shown by the works of Bai and collaborators \citep{Bai11,Bai14,Bai15}, and several other groups \citep{SS02b,W07,A11,WS12,LKF14,BLF16}. Another noteworthy result is that our system evolves towards a non-equipartition state. Moreover, we show that different choices of initial conditions for the kinetic and magnetic energies lead to the \emph{same} asymptotic state, for a \emph{fixed} Hall parameter. On the other hand, if the initial conditions are held fixed and the Hall parameter is varied, the final state is correspondingly altered. Thus, we demonstrate that the asymptotic state is sensitive to the values of the Hall parameter. 

Having outlined the rationale for studying the dynamics of Hall MRI turbulent stresses, we shall outline the contents of this paper. We present a preliminary description of Hall MHD and derive the governing equations for the turbulent stresses in Section \ref{HMHDPrimer}. This is followed by the presentation, and justification, of the closure model in Section \ref{SecClosure}. The ramifications of this model are explored in Section \ref{SecCSCase}. Finally, we conclude in Section \ref{SecConc}. A discussion of the Kazantsev small scale dynamo for Hall and extended MHD is presented in Appendix \ref{AppA}. We analyse the non-linear Alfv\'en wave solutions of Hall MHD, and their consequences for equipartition, in Appendix \ref{AppB}.

\section{Hall MHD: the mathematical preliminaries} \label{HMHDPrimer}
In this section, we provide a brief discussion of Hall MHD, and proceed to derive the small and large scale equations for Hall MHD.

\subsection{Basics of Hall MHD} \label{SSecBasicHMHD}
The equations of incompressible Hall MHD, when expressed in Alfv\'enic units, in a rotating frame are
\begin{eqnarray} \label{MomEqn}
\left[\frac{\p}{\p t} + {\bf v} \cdot \nabla\right] {\bf v} + 2 \boldsymbol{\Omega} \times {\bf v} &=& \left(\nabla \times {\bf b}\right) \times {\bf b} + \nu \Delta {\bf v} \\
&&- \nabla \left(\frac{p}{\rh} + \Phi - |\boldsymbol{\Omega} \times {\bf r}|^2\right) \nonumber,
\end{eqnarray}
\begin{equation} \label{OhmLaw}
\left[\frac{\p}{\p t} + {\bf v} \cdot \nabla\right] {\bf b} = \left({\bf b}\cdot \nabla\right) {\bf v} - d_i \nabla \times \left[\left(\nabla \times {\bf b}\right) \times {\bf b}\right] + \eta \Delta {\bf b},
\end{equation}
where ${\bf b} = {\bf B}/\sqrt{\mu_0 \rh}$ is the Alfv\'en velocity, $d_i$ is the ion skin depth (in SI units), and $\Delta$ represents the Laplacian operator. Firstly, we observe that the momentum equation (\ref{MomEqn}) is identical to that of ideal MHD. Secondly, in the limit $d_i \rightarrow 0$, we see that (\ref{OhmLaw}) reduces to the ideal MHD Ohm's law. The limit $d_i \rightarrow 0$ is not entirely appropriate as we take the vanishing limit of a  dimensional quantity; it is more appropriate to take the limit $\lambda_H:=d_i/L \rightarrow 0$, where $\lambda_H$ is the Hall parameter that measures the ratio of the ion skin depth to the scale length of the system $\left(L\right)$. In subsequent discussions, we shall assume that this limit is taken implicitly, although we use the notation $d_i \rightarrow 0$. However, a cautionary word is necessary, since Hall MHD is a \emph{singular} perturbation of ideal MHD \citep{YM99}, and a naive application of this limit may sometimes lead to misleading results. 

Before proceeding further, a comment regarding $d_i$ is in order. For a fully ionized plasma, it is known that $d_i = 2.27 \times 10^7 n^{-1/2}$ cm. From this expression, it is evident that $d_i$ is typically ``small'' in astrophysical systems, even when it is compared against the inertial range of the turbulence, as the latter serves as the domain of the small-scale dynamo. However, there are two potential pathways by which the Hall parameter (or its equivalent) can be rendered larger:
\begin{itemize}
\item Many astrophysical plasmas, such as protoplanetary discs, are \emph{not} fully ionized. The presence of neutrals alters the dynamics considerably. The incorporation of neutrals leads to a model with the same structure as Hall MHD but the ion skin depth $d_i$ is replaced by $d_i \sqrt{\rh_n/\rh_i}$, where $\rh_n$ and $\rh_i$ are the mass densities of the neutrals and ions respectively \citep{KL13}. Since $\sqrt{\rh_n/\rh_i} \gg 1$ in protoplanetary discs, one can obtain a modified (ion) skin depth of approximately $0.1$ AU \citep{W07,KL13,LKF14}, which is much larger than its fully ionized counterpart.
\item For a three species plasma consisting of electrons, ions and singly charged dust particles, the structure is again identical to Hall MHD. However, it has been shown that the ion skin depth is replaced by its dusty counterpart $\lambda_d = c/\omega_{pd}$, which is typically much larger. As a result, this can once again lead to a higher value of the (dust) Hall parameter \citep{MLin15}.
\end{itemize}
Thus, we can conclude that the presence of non-ideal MHD effects such as neutrals and/or dust is likely to give rise to a higher (effective) Hall parameter. As a result, our analysis is likely to be of interest in systems where such effects play an important role. 

\subsection{The Hall MHD small and large scale equations}
In our subsequent discussion, we shall use ${\bf v} = {\bf V}_0 + \hat{\bf v}$ and ${\bf b} = {\bf B}_0 + \hat{\bf b}$ where the upper case characters and the subscript `$0$' denote the mean fields, whilst the overhat denotes the fluctuating contributions. The mean-field equations are
\begin{eqnarray} \label{MFVel}
&& \left(\p_t + V_{0j} \p_j \right) V_{0i} + 2 \epsilon_{ijk} \Omega_j V_{0k} \nonumber \\
&&\, = - \p_i \Pi_0 + B_{0j} \p_j B_{0i} + \nu \Delta V_{0i} + \p_j \left(\langle{{M}_{ij} - {R}_{ij}}\rangle\right),
\end{eqnarray}
\begin{eqnarray} \label{MFB}
&& \left(\p_t + V_{0j} \p_j \right) B_{0i} = B_{0j} \p_j V_{0i} + \eta \Delta B_{0i} + \p_j \langle{\calf_{ij}}\rangle \nonumber \\
&& \quad \quad \quad \quad \quad \quad \quad - d_i B_{0j} \p_j \calj_{0i} + d_i \calj_{0j} \p_j B_{0i},
\end{eqnarray}
where we introduce the definitions
\begin{eqnarray}
&& \Pi = \frac{p}{\rh} + \frac{b^2}{2} + \Phi - |\boldsymbol{\Omega} \times {\bf r}|^2,\nonumber \\
&& \boldsymbol{\calj} = \nabla \times {\bf b}, \nonumber \\
&& {\bf u}^{(e)} = {\bf v} - d_i \boldsymbol{\calj},
\end{eqnarray}
and we also note that $\p_j V_{0j} = 0 = \p_j B_{0j}$. For (\ref{MFVel}) and (\ref{MFB}) to be closed, we require knowledge of the evolution of
\begin{equation} \label{MaxStress}
M_{ij} = \hat{b}_i \hat{b}_j,
\end{equation}
\begin{equation} \label{ReyStress}
R_{ij} = \hat{v}_i \hat{v}_j,
\end{equation}
\begin{equation} \label{FarStress}
\calf_{ij} = \hat{u}^{(e)}_i \hat{b}_j - \hat{u}^{(e)}_j \hat{b}_i,
\end{equation}
representing the Maxwell, Reynolds and modified Faraday stresses respectively. In the limit $d_i \rightarrow 0$, we see that $\calf_{ij}$ reduces to the usual Faraday stress tensor
\begin{equation}
F_{ij} = \hat{u}_i \hat{b}_j - \hat{u}_j \hat{b}_i.
\end{equation}
In our subsequent discussion, we shall assume that ${\bf B}_0 = 0 = \langle{\calf_{ij}\rangle}$, thereby eliminating (\ref{MFB}) altogether. As a result, we only require the evolution equations for (\ref{MaxStress}) and (\ref{ReyStress}) respectively. This constraint is tantamount to assuming that there is no mean magnetic field, and the large-scale dynamo is entirely absent; its presence would lead to the existence of finite contributions to the electromotive force. It is evident, of course, that such a prescription is likely to be unphysical, given the concomitant existence of large scale fields and turbulent stresses, but our approach is motivated by the earlier strategies adopted by \cite{KY93,KY95,GIO03,PCP06,PCP08}.

With these simplifications, the induction equation for the fluctuating component of the magnetic field is
\begin{eqnarray}
\left(\p_t + V_{0k} \p_k \right) \hat{b}_i = \hat{b}_k \p_k V_{0i} + \p_k \calf_{ik} + \eta \Delta \hat{b}_i,
\end{eqnarray}
and we can use this to obtain the governing equation for the Maxwell stress as follows:
\begin{eqnarray}
&& \left(\p_t + V_{0k} \p_k \right) \langle{M_{ij}}\rangle - \langle{M_{ik}}\rangle \p_k V_{0j} - \langle{M_{jk}}\rangle \p_k V_{0i} \nonumber \\
&& \quad = \langle{\hat{b}_j \p_k \calf_{ik} + \hat{b}_i \p_k \calf_{jk}}\rangle + \eta \langle{\hat{b}_j \Delta \hat{b}_i + \hat{b}_i \Delta \hat{b}_j}\rangle.
\end{eqnarray}
This can be re-expressed in terms of $F_{ij}$, yielding
\begin{eqnarray} \label{MaxwellEvol}
&& \left(\p_t + V_{0k} \p_k \right) \langle{M_{ij}}\rangle - \langle{M_{ik}}\rangle \p_k V_{0j} - \langle{M_{jk}}\rangle \p_k V_{0i} \nonumber \\
&& \quad = \langle{\hat{b}_j \p_k F_{ik} + \hat{b}_i \p_k F_{jk}}\rangle + \eta \langle{\hat{b}_j \Delta \hat{b}_i + \hat{b}_i \Delta \hat{b}_j}\rangle \nonumber \\
&& \,\,\, - d_i \Big\langle{\hat{b}_j \p_k \left(\hat{\calj}_i \hat{b}_k - \hat{\calj}_k \hat{b}_i\right) + \hat{b}_i \p_k \left(\hat{\calj}_j \hat{b}_k - \hat{\calj}_k \hat{b}_j\right)}\Big\rangle,
\end{eqnarray}
and it is evident that the last line vanishes upon $d_i \rightarrow 0$, and the ensuing expression coincides with the result obtained in \citet{GIO03}. We see that the last line is comprised of three-point correlation functions, each of which involves terms that are cubic in $\hat{b}$ and its derivatives. We shall return to these terms shortly hereafter; for now, we proceed by applying the same approach to the fluctuating component of the equation of motion (\ref{MomEqn}).

By carrying out the same procedure, we find that the Reynolds stress evolves as per the equation:
\begin{eqnarray} \label{ReynoldsEvol}
&& \left(\p_t + V_{0k} \p_k \right) \langle{R_{ij}}\rangle + \langle{R_{ik}}\rangle \p_k V_{0j} + \langle{R_{jk}}\rangle \p_k V_{0i} \nonumber \\
&& + 2\epsilon_{jkl} \Omega_k \langle{R_{il}}\rangle + 2\epsilon_{ikl} \Omega_k \langle{R_{jl}}\rangle \nonumber \\
&& \quad = \nu \langle{\hat{v}_j \Delta \hat{v}_i + \hat{v}_i \Delta \hat{v}_j}\rangle - \langle{\hat{v}_i\p_j \widehat{\Pi} + \hat{v}_j\p_i \widehat{\Pi} }\rangle \nonumber \\
&& \quad + \langle{\hat{v}_i \p_k \left(M_{jk} - R_{jk}\right) + \hat{v}_j \p_k \left(M_{ik} - R_{ik}\right) }\rangle,
\end{eqnarray}
and it is found that this is identical to the expression derived in \citet{GIO03}. This is not particularly surprising as it's known that Hall MHD and ideal MHD possess the same equation of motion, but differing induction equations. 

From the next section onwards, we shall drop the symbol $\langle{\,}\rangle$, and it must be understood that all quantities (implicitly) have been subject to an ensemble averaging.

\section{Constructing a closure model}\label{SecClosure}
We will adopt the same principles described in Section 3.4 of \citet{GIO03}. The chief amongst them include $R_{ij} = 0 = M_{ij}$ and $M_{ij} = 0$ but $R_{ij} \neq 0$ constituting valid solutions (the latter is the hydrodynamic case); covariance of the model equations; positive semidefinite nature of the stresses being preserved. Moreover, it is also assumed that the non-linear terms do not exhibit any dependence on $\nabla {\bf V}$ or $\boldsymbol{\Omega}$, and that the viscosity and resistivity effects are neglected.\footnote{More precisely, \citet{GIO03} assumed that the model exhibits an ``asymptotic independence of $\mathrm{Re}$ and $\mathrm{Pm}$ in the limit $\mathrm{Re} \rightarrow \infty$.'' We note that this approximation is one of the model's limitations, as it cannot account for the dependence of the stresses on $\mathrm{Re}$ and $\mathrm{Pm}$ -- a factor of paramount importance in dynamo theory; see e.g. \citet{BS05,CH09} and references therein.} Lastly, it is dimensional considerations that impose key constraints on the form of the nonlinear terms, thereby leading \citet{GIO03} to his model. 

Firstly, we consider the limit $d_i \rightarrow 0$ and adopt the above assumptions, which leads us to the same two expressions that \citet{GIO03} arrives at:
\begin{eqnarray} \label{ReynoldsCSIdeal}
&& \left(\p_t + V_{k} \p_k \right) R_{ij} + R_{ik} \p_k V_{j} + R_{jk} \p_k V_{i} + 2\epsilon_{jkl} \Omega_k R_{il} \nonumber \\
&& + 2\epsilon_{ikl} \Omega_k R_{jl} = -\frac{C_1}{L} R^{1/2} R_{ij} - \frac{C_2}{L} R^{1/2} \left(R_{ij} - \frac{1}{3} R \delta_{ij} \right) \nonumber \\
&& \hspace{0.8 in} +\, \frac{C_3}{L}M^{1/2} M_{ij} - \frac{C_4}{L} M R^{-1/2} R_{ij},
\end{eqnarray}
\begin{eqnarray} \label{MaxwellCSIdeal}
&& \left(\p_t + V_{k} \p_k \right) M_{ij} - M_{ik} \p_k V_{j} - M_{jk} \p_k V_{i} \nonumber \\
&& \hspace{0.3 in} = \frac{C_4}{L} M R^{-1/2} R_{ij} - \frac{C_3 + C_5}{L}M^{1/2} M_{ij}.
\end{eqnarray}
In the above expressions, note that $R = \mathrm{Tr}\left(R_{ij}\right)$ and $M = \mathrm{Tr}\left(M_{ij}\right)$, whilst $L$ is the system scale length introduced in Sec. \ref{SSecBasicHMHD}, and the $C$'s are appropriate dimensionless coefficients. Moreover, we have dropped the averaging $\langle{\,}\rangle$ on $R_{ij}$ and $M_{ij}$, and also the subscript `$0$' on the mean-field velocity, to simplify the notation henceforth.

Now, we need to study how the Hall effect will affect this closure scheme. Upon a careful inspection of (\ref{MaxwellEvol}), we find that the only extra terms are those proportional to $d_i$. Moreover, each of these involves three factors of $\hat{b}$, and two spatial derivatives. Thus, our problem reduces to finding a simple closure scheme for third order correlation functions in terms of the second order ones. With this in mind, we propose that the new terms should be solely dependent on the tensors $M_{ij}$ and $\delta_{ij}$, with factors of $M^{1/2}$ and $M^{3/2}$ preceding them respectively. This is motivated by the fact that the Hall term in (\ref{OhmLaw}) does not feature any velocity-dependence. Moreover, the three-point correlations involving $d_i$ in (\ref{MaxwellEvol}) involve only factors of $\hat{b}$ and we have supposed that these terms can be written solely in terms on two-point correlations involving $\hat{b}$ exclusively, viz. $M$ and $M_{ij}$.

By employing this assumption and simple scaling principles, we modify (\ref{MaxwellCSIdeal}) to incorporate these new terms. Our revised evolution equation for $M_{ij}$ is given by
\begin{eqnarray} \label{MaxwellCS}
&& \left(\p_t + V_{k} \p_k \right) M_{ij} - M_{ik} \p_k V_{j} - M_{jk} \p_k V_{i}  \\
&& \hspace{0.2 in} = \frac{C_4}{L} M R^{-1/2} R_{ij} - \frac{C_3 + C_5}{L}M^{1/2} M_{ij} \nonumber \\
&& \hspace{0.2 in} - \lambda_H \left[\frac{C_6}{L} M^{1/2} M_{ij} + \frac{C_7}{L} M^{1/2} \left(M_{ij} - \frac{1}{3} M \delta_{ij} \right) \right], \nonumber
\end{eqnarray}
where $\lambda_H = d_i/L$ is the Hall parameter introduced earlier. To understand the origin of the Hall parameter in the above expression, consider the following heuristic argument. 

The last two terms on the RHS of (\ref{MaxwellEvol}) encapsulate the contributions from the Hall term. An inspection of these terms reveals the exist of two spatial gradients, and through dimensional considerations, these must be replaced by the corresponding inverse length scale, which is assumed to be $1/L$ for our model. Since there are two such factors, one of them combines with $d_i$ and yields $\lambda_H$. The other is visibly manifest as the $1/L$ factor appearing in the last two terms on the RHS of (\ref{MaxwellCS}). It must be noted that our model entails the implicit assumption $L \gtrsim d_i$. If one considers the case $L < d_i$, especially the regime $L \ll d_i$, this corresponds to the domain of magnetic reconnection \citep{Bisk00} with the formation of current sheets, etc. Although turbulent magnetic reconnection is an important area of research, it is not the object of study in this paper, and we shall restrict ourselves to astrophysical systems that obey $\lambda_H \lesssim 1$.

Upon taking the limit $\lambda_H \rightarrow 0$ in (\ref{MaxwellCS}), we recover (\ref{MaxwellCSIdeal}) as mandated. Moreover, we note that this new model preserves the principles outlined in Sec 3.4 of \citet{GIO03} as well. The most crucial inclusion of Hall MHD is that the uniscale problem is now rendered a biscale one, thanks to the presence of the ion skin depth. In order to further justify the two additional terms, we provide a brief physical interpretation of each of them. 

An examination of the terms involving $C_3$ (as well as $C_5$) and $C_6$ reveals that they have the same form. In the model of \citet{GIO03}, $C_3$ quantified the effects of the Lorentz force in the equation of motion (\ref{MomEqn}) which drives the turbulence. On the other hand, we find that a term akin to the Lorentz force is \emph{also} existent in the Ohm's law, as seen from the second term on the RHS of (\ref{OhmLaw}) which also involves the ion skin depth $d_i$. Thus, it is plausible to suppose that an additional term, akin to the one featuring $C_3$, but involving the skin depth, emerges. This additional term is manifest as $C_6$ in (\ref{MaxwellCS}). Next, let us consider $C_7$ in  (\ref{MaxwellCS}) which clearly represents a drive towards (or away from) isotropization. Since there is no \emph{a priori} reason from observations or simulations to suppose that Hall MHD supports enhanced (or reduced) isotropization, we anticipate that $C_7$ will typically end up being zero. In our subsequent analysis, we shall mostly operate with this assumption $\left(C_7 = 0\right)$, but it is important to recognize that this issue can be settled decisively only through future numerical simulations of Hall MRI turbulent stresses.

We have specified an ansatz for the evolution of $M_{ij}$ in 
(\ref{MaxwellCS}), but we have not commented hitherto on how $R_{ij}$ behaves. Given that (\ref{ReynoldsEvol}) is identical to that of ideal MHD, it is tempting to conclude that (\ref{ReynoldsCSIdeal}) holds true even for this model. However, an important clue emerges from the energy equation. Let us take the trace of  (\ref{MaxwellCS}) and (\ref{ReynoldsCSIdeal}), and sum the two equations. The end result is
\begin{eqnarray} \label{EngDiss}
\left(\p_t + V_k \p_k\right) \left(\frac{R}{2}+\frac{M}{2}\right) &=& \left(M_{ij} - R_{ij}\right) \p_j V_i - \frac{C_1}{2L}R^{3/2} \nonumber \\
&&- \frac{C_5}{2L}M^{3/2} - \lambda_H \frac{C_6}{2L}M^{3/2}.
\end{eqnarray}
A comparison with \citet{GIO03} reveals that only the last term on the RHS is different. However, a crucial aspect of the Hall MHD must be stated here - it is well known that the total energy of Hall MHD is \emph{identical} to that of ideal MHD in the limit of vanishing viscosity and resistivity \citep{LMM15}. Thus, this would imply that the RHS of (\ref{EngDiss}) and equation (30) of \citet{GIO03} must equal one another, since the LHS of the two expressions constitute the same energy. In turn, this implies that we cannot use (\ref{ReynoldsCSIdeal}) directly. Instead, we must include the $C_6$ term from (\ref{MaxwellCS}), albeit with an opposite sign. Thus, the evolution equation for the Reynolds stress is given by
\begin{eqnarray} \label{ReynoldsCS}
&& \left(\p_t + V_{k} \p_k \right) R_{ij} + R_{ik} \p_k V_{j} + R_{jk} \p_k V_{i} + 2\epsilon_{jkl} \Omega_k R_{il} \nonumber \\
&& + 2\epsilon_{ikl} \Omega_k R_{jl} = -\frac{C_1}{L} R^{1/2} R_{ij} - \frac{C_2}{L} R^{1/2} \left(R_{ij} - \frac{1}{3} R \delta_{ij} \right) \nonumber \\
&& \hspace{0.8 in} +\, \frac{C_3}{L}M^{1/2} M_{ij} - \frac{C_4}{L} M R^{-1/2} R_{ij} \nonumber \\
&& \hspace{0.8 in} +\, \lambda_H \frac{C_6}{L} M^{1/2} M_{ij}.
\end{eqnarray}

\begin{figure*}
$$
\begin{array}{ccc}
 \includegraphics[width=5.4cm]{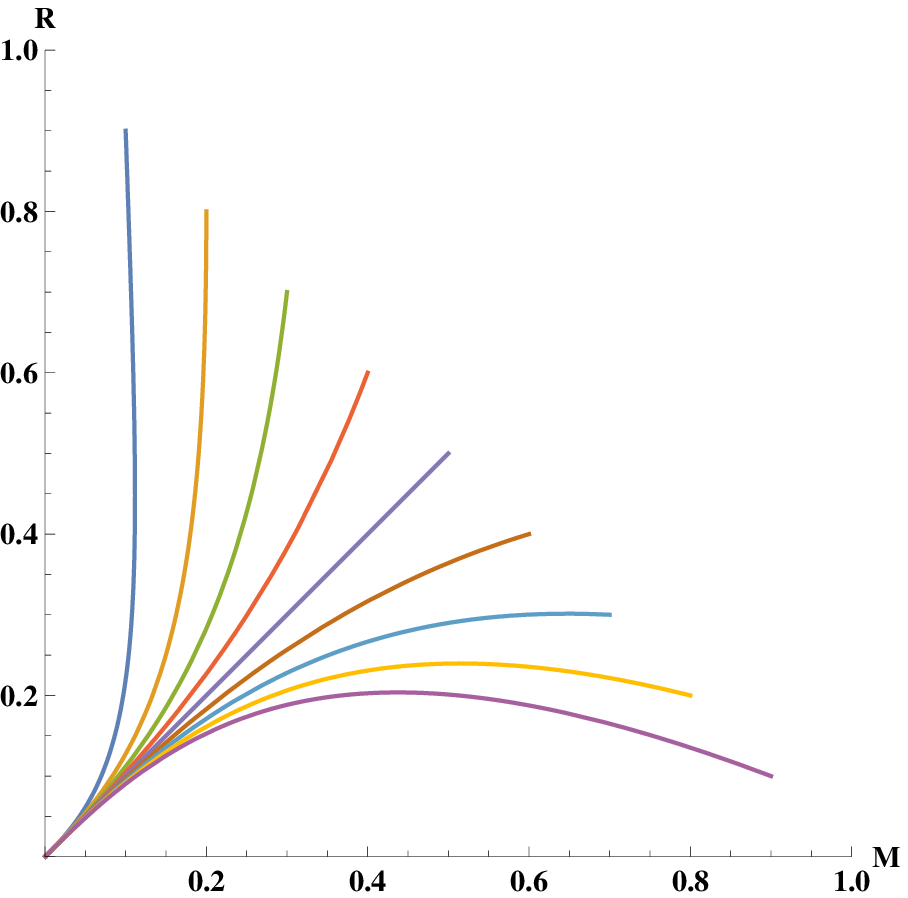} & \includegraphics[width=5.4cm]{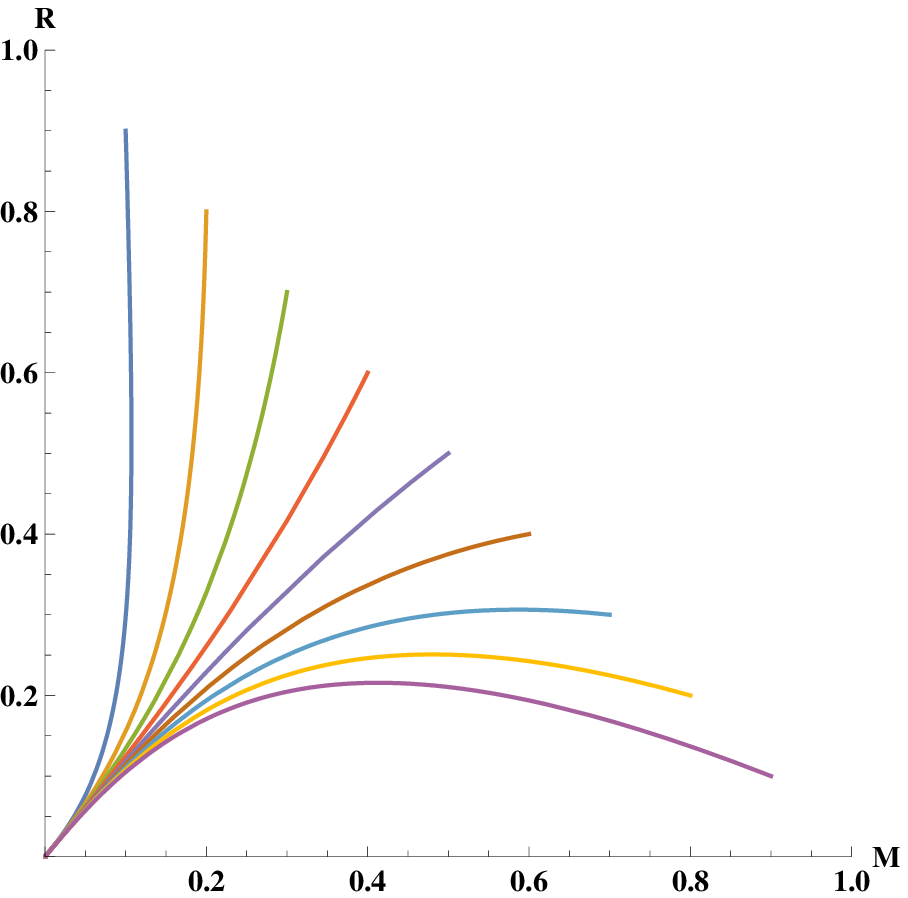} & \includegraphics[width=5.4cm]{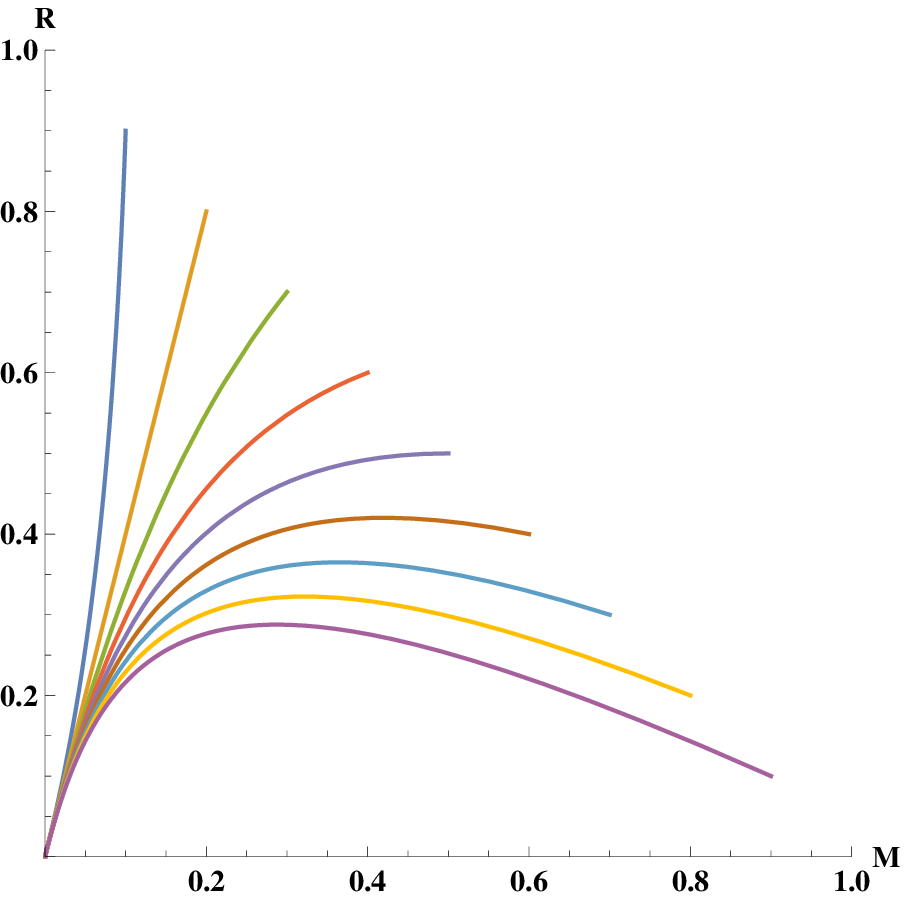}\\
 \quad\quad(a) & \quad\quad(b) & \quad\quad(c)\\
\end{array}
$$
\caption{(colour figures online) The phase diagram of $R$ vs $M$ is presented for the three figures. Panels (a), (b) and (c) correspond to $\lambda_H = 0$, $\lambda_H = 0.1$ and $\lambda_H = 1$ respectively. }
\label{fig1}
\end{figure*}

\begin{figure*}
$$
\begin{array}{ccc}
 \includegraphics[width=5.4cm]{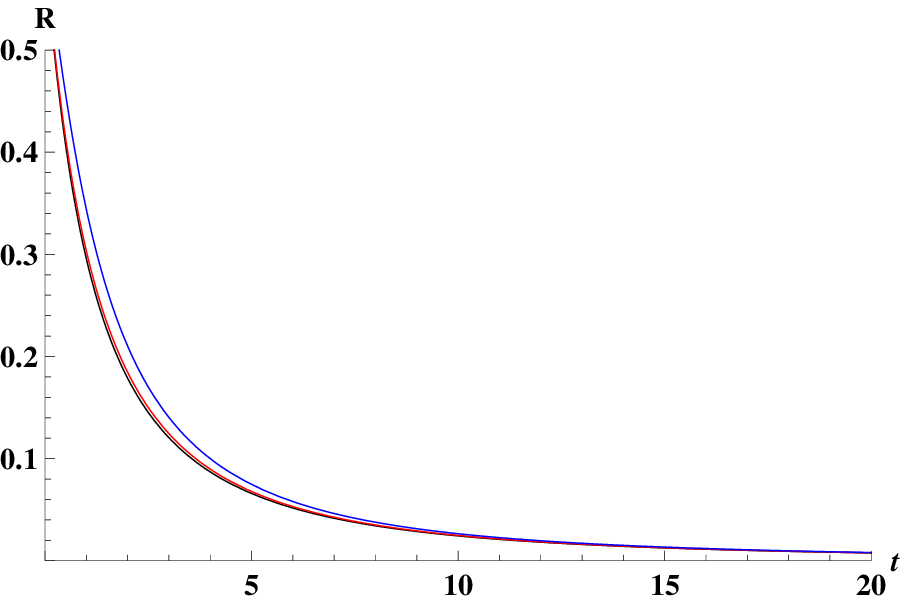} & \includegraphics[width=5.4cm]{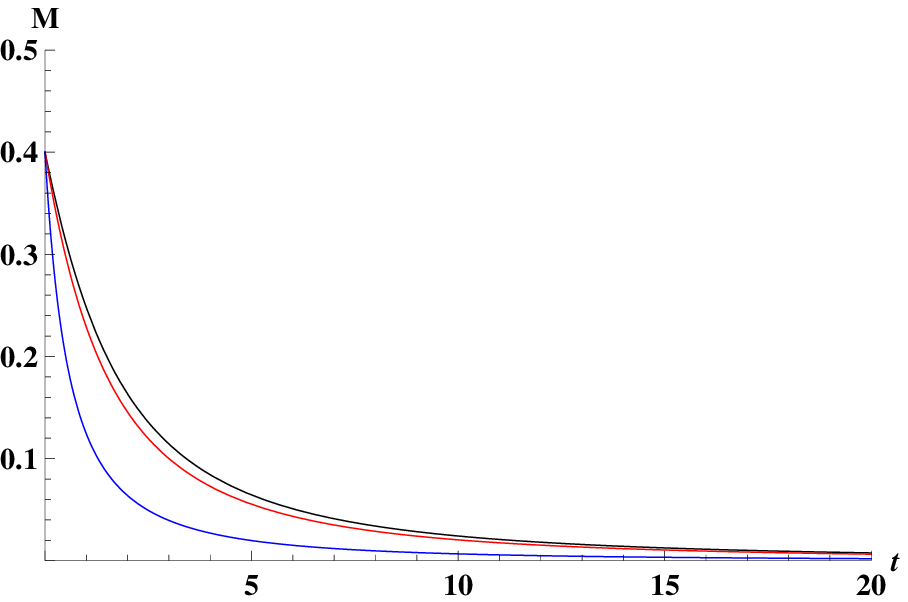} & \includegraphics[width=5.4cm]{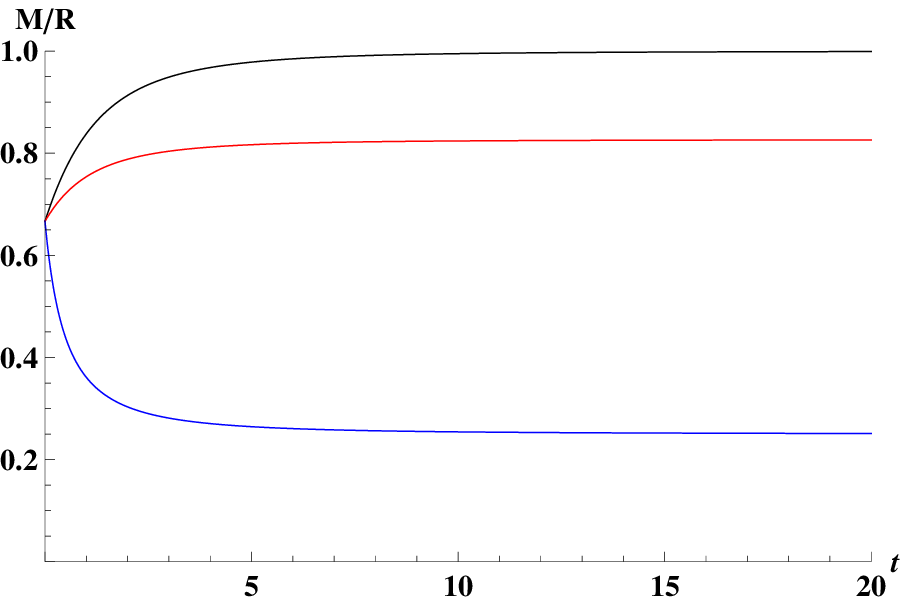}\\
 \quad\quad(a) & \quad\quad(b) & \quad\quad(c)\\
\end{array}
$$
\caption{(colour figures online) Panels (a) and (b) depict the turbulent kinetic and magnetic energies, whilst Panel (c) determines the ratio of the magnetic and the kinetic energies. In each of these panels, note that the black, red and blue curves correspond to $\lambda_H = 0$, $\lambda_H = 0.1$ and $\lambda_H = 1$ respectively. In each of these panels, the initial conditions correspond to $R=0.6$ and $M=0.4$.}
\label{fig2}
\end{figure*}

However, we note that such a reasoning is partly fallacious, since one can start with an energy-conserving (non-dissipative) model and recover dissipative effects when studying turbulence - a result of the non-linearity and the averaging processes involved. For instance, when studying dynamo theory, it is well known that the electromotive force contributes a turbulent resistivity (by means of the $\beta$-effect), which is a dissipative process \citep{BS05}. Hence, we modify the $C_6$ term in (\ref{MaxwellCS}) to $C_6'= C_6 + C_8$, where $C_8 > 0$. This introduces an additional turbulent contribution to the energy, i.e. we end up with (\ref{EngDiss}) except that the coefficient of $C_6$ is replaced by $C_8$ instead. We conclude by noting that (\ref{MaxwellCS}) with $C_6'$ and (\ref{ReynoldsCS}) shall serve as our (minimal) model equations for Hall MHD turbulent stresses. We demand that $C_6 \neq 0$, but we allow for the possibility that $C_7 = 0 = C_8$. 

Before concluding this Section, let us summarize our model and its governing equations, for the purpose of clarity. For a \emph{fully} self-consistent treatment of the MRI, a model that co-evolves the large scale fields $V_0$ and $B_0$ and the turbulent stresses $M_{ij}$, $R_{ij}$, and $\mathcal{F}_{ij}$ is required. The need for a fully consistent model along these lines has been emphasized in \citet{Bl12,BN15} and a similar program was implemented in \citet{SB15} recently. In this paper, we have chosen to focus primarily on the stresses, and we chose to artificially shut off the large scale magnetic field (along with $\mathcal{F}_{ij}$) for the sake of simplicity. Thus, our reduced model is fully specified by (\ref{MaxwellCS}) with $C_6'$ and (\ref{ReynoldsCS}), in conjunction with the mean field equation (\ref{MFVel}); in the latter, it must be noted that $B_0 = 0$ on account of dropping the mean magnetic field from the system.

In the next Section, we shall further restrict ourselves to the case where the (large scale) velocity $V$ is \emph{externally} specified in (\ref{MaxwellCS}) and (\ref{ReynoldsCS}), and is not consistently co-evolved by means of (\ref{MFVel}). This situation is very similar, although not equivalent, to kinematic dynamo theory, where the velocity field in the induction equation is prescribed externally. Secondly, we shall study the case where the stresses are homogeneous, viz. we drop terms of the form $\p_k M_{ij}$ and $\p_k R_{ij}$. Each of these assumptions are \emph{not} a fundamental limitation of our model - one can revoke these assumptions, and study the full equations noted above instead. We offer a further comment on our approach in Appendix \ref{AppA3}

The primary reasons for enforcing these two assumptions are twofold. Firstly, it enables a straightforward comparison with previous models of MRI stresses, albeit in the context of ideal MHD.  For instance, it is worth mentioning that the central results of \citet{KY93,KY95}, \citet{GIO03} and \citet{PCP06} are presented in a similar form, and this enabled \citet{PCP08} to undertake a direct comparison of these models alongside the $\alpha$-viscosity prescription. The above two assumptions make the comparison with these earlier models more transparent, as our equations are now akin to these efforts. Secondly, as this serves as our first foray into building a heuristic model of the stresses, we wish to extract qualitative features of the model. As we shall show subsequently, we do witness the emergence of distinct differences between the Hall and ideal MHD cases. 

\section{Specific cases of the closure scheme} \label{SecCSCase}

Firstly, let us observe that taking $M_{ij} \rightarrow 0$ ensures that (\ref{MaxwellCS}) vanishes identically and (\ref{ReynoldsCS}) reduces to the hydrodynamic stress equation of \citet{GIO03}, thereby preserving one of the postulates of that model. Secondly, let us consider the case where $R_{ij} \rightarrow 0$ and ignore shear-effects, rotation, and the spatial dependence of $M_{ij}$. We end up with
\begin{equation}
\p_t M = - \left(\frac{C_5 + \lambda_H C_8}{L}\right) M^{3/2},
\end{equation}
which yields $M = 4L^2 \left(C_5 + \lambda_H C_8\right)^{-2} t^{-2}$ upon solving. If $\lambda_H \rightarrow 0$, we see that the MHD result follows, but it is interesting to consider the effects of a finite $\lambda_H$ - we find that $M$ is \emph{always} smaller than its MHD value (with $\lambda_H = 0$) as long as $\lambda_H C_8$ and $C_5$ are of the same sign. It suggests that the Hall term leads to a reduction in the turbulent magnetic energy in contrast to the MHD case; moreover, the variation of $M$ with $\lambda_H$ is also monotonic. 

Next, let us consider the case with no rotation and shear, and the stresses are taken to be spatially homogeneous. The turbulent energies are given by
\begin{equation}
\p_t R = \frac{-C_1 R^{3/2} + \left(C_3 + \lambda_H C_6 \right) M^{3/2} - C_4 R^{1/2} M}{L} ,
\end{equation}
\begin{equation}
\p_t M = \frac{C_4 R^{1/2} M - \left(C_3 + \lambda_H C_6 + C_5 + \lambda_H C_8 \right) M^{3/2} }{L},
\end{equation}
and we observe that these equations are exactly identical to those of \citet{GIO03} provided that we introduce the new variables $C_3' = C_3 + \lambda_H C_6$ and $C_5' = C_5 + \lambda_H C_8$. As a result, if we choose all the $C$'s to be unity, we find that the system is driven towards a final state wherein $M/R = \left(C_4/C_3'\right)^2 = \left(1+\lambda_H\right)^{-2}$. We find that a very interesting mode of behaviour emerges -- the asymptotic value of the ratio $M/R$ declines from unity for $\lambda_H \rightarrow 0$ to attain a minimum value of $1/4$ at $\lambda_H = 1$. In other words, we find that $M/R < 1$ for all values of $\lambda_H > 0$, and this ratio declines monotonically as $\lambda_H$ is increased.

In other words, we can conclude that the Hall effect results in a scenario wherein the turbulent magnetic energy is lesser than its kinetic counterpart, for our choice of coefficients and with $\lambda_H > 0$. The lack of equipartition is consistent with numerical studies of large-scale Hall MHD dynamos, see e.g. \citet{MGM03b}. However, it must be emphasized that the setup (and physics) for these dynamo simulations was quite different from the analysis undertaken herein. We also note that the solar wind has been shown to exhibit non-equipartition values, although the opposite result, viz. $M/R > 1$, is typically observed \citep{WBP11,CBSM13}. In order to recover this result $\left(M/R > 1\right)$, one can tune the parameters of our phenomenological model accordingly, as illustrated below through an example.

We reiterate the central assumption invoked in obtaining our results, namely that all the coefficients were taken to be \emph{unity}.  Suppose that we consider a case where $C_3 = C_6$ and $\left(C_4/C_3\right)^2 = 2$; the chosen values for these parameters do not violate any of the constraints of our model. We end up with $M/R = 2 \left(1+\lambda_H\right)^{-2}$. Interestingly, we find that $M/R = 2$ for $\lambda_H = 0$ and $M/R = 1/2$ for $\lambda_H = 1$. The result indicates that the model can lead to both $M/R < 1$ and $M/R > 1$ for different values of the Hall parameter but for the same ratio of $C_4/C_3$. In this example, the transition occurs at $\lambda_H = 0.414$. 

At this juncture, we also wish to point out another important fact. We have noted that $M/R = \left(C_4/C_3'\right)^2$ in Hall MHD, whilst $M/R = \left(C_4/C_3\right)^2$ in ideal MHD. It is evident that the two ratios will coincide if and only if $C_3' = C_3$, viz. when $\lambda_H = 0$.  Thus, if the conditions $C_5 < 0$ and $C_3,\,C_3' > 0$ are satisfied, it is easy to verify that $M/R$ will be higher in the Hall MHD case. A recent study by \citet{BLF16} is striking in this regard, as it demonstrates, by means of numerical simulations, that $M/R$ is \emph{different} for ideal and Hall MHD, and that the latter exhibits an enhanced value of $M$ owing to the Hall effect. On account of the free parameters present in our model, we see that the simulations of \citet{BLF16} (in particular, their Fig. 5) are entirely consistent with our conclusions. This result does not constitute a `proof' of the veracity of our model, but it does lend it additional credence. In Appendix \ref{AppB}, we have also studied (nonlinear) Alfv\'en waves, an important component of understanding plasma turbulence, in Hall MHD. We argue therein that the lack of equipartition of the wave energy is consistent with our results, and may indicate that this phenomenon may be endowed with some degree of universality.

In Fig. \ref{fig1}, the phase diagram, i.e. the plot of $R$ vs $M$ for different choices of initial conditions is shown. When we consider panel (a) in Fig. \ref{fig1}, we find that there is a drive towards equipartition for $\lambda_H = 0$, where all the $C$'s have been chosen to be unity. Moreover, the system is driven towards the fixed point $R = M = 0$. Both of these results correspond to the ideal MHD case and agree with the findings of \citet{GIO03}. On the other hand, panel (c) clearly illustrates that there is a drive towards the steady state $R = M = 0$ but the path taken to achieve this state does \emph{not} correspond to equipartition. We also find that holding the Hall parameter \emph{fixed} leads to the same asymptotic state for different choices of the initial conditions. However, if we hold the initial conditions and the coefficients ($C_3$, $C_4$ and $C_6$) \emph{fixed}, but the Hall parameter is \emph{varied}, the asymptotic state reached by the system is duly affected. Our findings indicate that the asymptotic state of the system is regulated by the Hall parameter (and the coefficients inherent in the model), but not the initial conditions. 

From Fig. \ref{fig2}, we find that the kinetic energy \emph{increases slightly} when we increase the value of $\lambda_H$. On the other hand, the magnetic energy \emph{clearly decreases} when we increase $\lambda_H$. From panel (c) of Fig. \ref{fig2}, we see that the ratio of $M/R$ starts off identically for each of the three cases, but evolves to a very different final value; for $\lambda_H = 0$, it corresponds to equipartition whilst $\lambda_H = 1$ yields $M/R = 0.25$, in agreement with the preceding discussion. We also note an important result that is evident from panel (c) - for the same initial value of $M/R$, the asymptotic value of $M/R$ reached by the system can be either \emph{greater} or \emph{smaller} than this initial value, and the increase/decrease is regulated by the value of $\lambda_H$. Lastly, note that our model is `symmetric' about $\lambda_H = 1$ and hence we note that $\lambda_H = 0$ and $\lambda_H = \infty$ are equivalent, as are $\lambda_H = 0.1$ and $\lambda_H = 10$. 

Hitherto, we have ignored both the rotation and shear for the sake of simplicity. This amounted to setting $V_k = 0 = \Omega_k$ in (\ref{MaxwellCS}) and (\ref{ReynoldsCS}). Yet, it is widely known that the MRI is a fundamentally anisotropic, and shear-driven, phenomenon. Thus, a full investigation of this model would require the inclusion of the shear (and also the rotation). When we set $C_7 = 0$ and study the case with finite shear and rotation, we find that the resultant behaviour is akin to that of Fig. 5 in \citet{GIO03}. In other words, we do find that the stresses are rendered anisotropic, along the expected lines.

As the qualitative behaviour, with the Hall effects, is (mostly) identical to that of the ideal MHD model, we have not reproduced the figures here. To be more precise, we find that the general shape and structure of the stresses is akin to Fig. 5 of \citet{GIO03}, and the only difference is an overall shift (upwards or downwards) that depends on the signs of $C_5$, $C_6$ and $C_8$. Thus, the presence of a finite $\lambda_H$ successfully enhances (or dampens) the saturated values of the stresses, depending on the exact magnitudes of the chosen coefficients. Broadly speaking, we find that the results first outlined in \citet{SS02b}, and verified later by several authors, are recovered, such as a positive Hall parameter giving rise to enhanced saturated stresses.

\section{Discussion and Conclusion} \label{SecConc}
In the Introduction, we have offered compelling reasons as to why the standard Shakura-Sunyaev $\alpha$ viscosity prescription fails to capture the subtleties of the MRI turbulent stresses. Moreover, we have also indicated the considerable importance of the Hall drift in a plethora of astrophysical settings. Motivated by these two facets, the aim of the paper was to extend the work of \citet{GIO03} on MRI turbulent stresses (in ideal MHD) to Hall MHD.

After summarizing the basics of Hall MHD, we elucidated the basic tenets of the \citet{GIO03} model. Subsequently, we carried out the algebraic manipulations of the Hall MHD equations, and implemented the closure scheme proposed by \citet{GIO03} in the context of ideal MHD. However, the Hall MHD terms gave rise to additional contributions, which are on the order of $\mathcal{O}\left(d_i/L\right)$. We modelled these contributions by appealing to dimensional arguments, the absence of any velocity field in the Hall term, and consistency with ideal MHD and the \citet{GIO03} model. It resulted in three new contributions which affected the backreaction of the Lorentz force, dissipation and isotropization. Of these, the latter was quantified by the coefficient $C_7$, which we believe can be set to zero. This stems from the fact that simulations of Hall MHD have not reported any particular drive towards, or away, from isotropization. If such results are discovered in the future, a suitable non-zero value of $C_7$ can be specified accordingly.

There were two chief results that emerged from our Hall MRI model. The first was the lack of equipartition, in general, between the magnetic and kinetic energies for a non-zero value of the Hall parameter. In fact, we find that the system approaches an asymptotic ratio of the energies that varies with the Hall parameter. Our findings are supported by the recent simulations of Hall MRI turbulence by \citet{BLF16}, who arrived at a similar result, as they concluded that the Hall term does affect the ratio of the turbulent magnetic and kinetic stresses (and energies). In Appendix \ref{AppB}, we have tackled this aspect from a different standpoint by examining nonlinear Alfv\'enic waves in Hall MHD, which are known to exhibit a lack of equipartition. Given the importance of Alfv\'en waves in plasma turbulence, we believe that the absence of equipartition may prove to be a fairly robust characteristic of Hall MHD.

There are astrophysical systems, such as the solar wind, which have been shown to exhibit this lack of equipartition \citep{WBP11,CBSM13}. Hence, our model could, perhaps, be employed in such a context to model the turbulent stresses and energies involved. Our surmise is quite reasonable since Hall MHD has been successfully employed to model turbulence in the solar wind in the past few decades \citep{GSRG96,GBuc07,ACVS08,BC13}. We have also shown that our model leads to an asymptotic ratio of the energies that is independent of the choice of initial conditions. On the other hand, it is sensitive to the model parameters $C_3$, $C_4$ and $C_6$ and the Hall parameter. The latter constitutes an important result since one can end up with vastly different ratios of the magnetic to kinetic energy even when all the other variables except for the Hall parameter are held fixed.

The second important finding is that the MRI stresses are rendered anisotropic when rotation and shear are taken into account, in agreement with physical intuition and the results presented in \citet{GIO03}. The MRI stresses are also duly modified when the Hall-induced terms are included. This has immediate consequences for protoplanetary discs, where the Hall term has been proposed as a prime candidate in reviving ``dead zones''. By including the effects of rotation and shear in our model, we are led to conclude that its findings are in basic agreement with the numerical simulations carried out by \citet{SS02b} as well as multiple groups in recent times \citep{Bai11,KL13,LKF14,Bai14,Bai15,SLKA15}. The coefficients of our model, when tuned appropriately, lead to enhanced values of the saturated MRI stresses for a finite Hall parameter (as opposed to the ideal MHD case).

Since there are three (or possibly two if $C_7 = 0$) contributions arising from the Hall term, we carried out a comparison against the well-known kinematic Kazantsev prescription, typically used in small-scale dynamos, in Appendix \ref{AppA}. Owing to the numerous assumptions employed in the Kazantsev model, such as homogeneity, isotropy and Gaussian statistics, we demonstrated that the Hall term does not yield any finite contributions. We believe that this constitutes a good example of the inherent limitations of the Kazantsev prescription. However, we also showed that the inclusion of electron inertia leads to non-trivial, albeit very `small', contributions when using the Kazantsev approach. Lastly, we have delineated the differences between our model and that of \citet{Kaz68} to avoid potential ambiguities.

As our model is fairly simple, it may not capture all the relevant physics unfolding in Hall MRI turbulence. Nevertheless, we have shown that it yields interesting results (the potential absence of equipartition, enhanced stresses), and that it is a better choice than the Kazantsev prescription which yields a trivial result. The Hall MRI model developed here can be `fitted' by carrying out numerical simulations and estimating the values of the free parameters $\left(C_1\, \,\mathrm{to}\, \,C_8\right)$ that characterize our model. If the `fit' proves to be successful, it is likely that our model could serve as an effective formalism for studying a host of astrophysical environments where MRI turbulence and the Hall term play a significant role.

As we have primarily concerned ourselves with investigating MRI stresses in the context of the $\alpha$ viscosity prescription, it was quite reasonable to ignore the mean magnetic field. Yet, one expects the preponderance of large-scale magnetic fields in many astrophysical objects \citep{BS05}, suggesting that a further treatment along these lines would be beneficial. In the past decade, \citet{MGM03b,MGM05} had already undertaken simulations of Hall MHD large-scale dynamos. One of the striking results therein was the lack of equipartition between the \emph{large scale} energies, rather analogous to our model where a similar result was obtained for the \emph{turbulent} energies. For this reason, we hope to generalize our model in the future to analytically predict, and explain, the conclusions of \citet{MGM03b,MGM05}.

Although the aforementioned studies of mean-field Hall MHD dynamos do exist, they were not explicitly concerned with modelling the Hall MRI dynamo. On this front, quite recently, \citet{KL13,BLF16} demonstrated, by means of numerical simulations, that the Hall effect reduced turbulent transport and gave rise to ``zonal'' structures. A subsequent study of stratified Hall MRI turbulence by \citet{LKF14} could generate a large-scale Maxwell stress, provided that the correct polarity, viz. $\boldsymbol{\Omega}\cdot{\bf B} > 0$, is prevalent; a similar set of results was also presented in \citet{SLKA15}. It is interesting to note that the ``zonal'' structures of \citet{KL13,BLF16} were not observed in \citet{LKF14}, possibly because the latter employed a stratified model.

It will be our goal in future investigations to construct a mean-field theory that accounts for the evolution of the stresses and the large-scale fields by extending the formalism outlined in this paper. Namely, we hope to couple the exact evolution equations to an appropriate closure scheme dictated by dimensional and physical considerations, analogous to the approach utilized here. We note that some semi-analytical models have already appeared in the literature \citep{KL13}, but it is our intention to design such models from scratch by means of the above procedure. A successful model could pave the way to understanding the exact role of the Hall term in the MRI dynamo. Furthermore, we anticipate that the model may play an important role in modelling neutron stars, protoplanetary discs, solar wind, and solar physics in general \citep{Mie05,HRW12}, as the Hall term is crucial in shaping the dynamics of these systems.

\section*{Acknowledgments}
ML and AB were supported by the DOE (Grant No. DE-AC02-09CH-11466) and the NSF (Grant No. AGS-1338944) during the course of this work. ML is grateful to Gordon Ogilvie for his encouraging comments regarding a preliminary version of the manuscript. 


\appendix

\section{A Kazantsev Extended MHD dynamo} \label{AppA}
In our design of the Hall MHD turbulent stresses, we have introduced two new coefficients $C_6$ and $C_8$, with the possibility of an extra isotropization term $C_7$. However, a simplified treatment leads to the total \emph{absence} of these extra terms in the Maxwell stress evolution equation (\ref{MaxwellCS}). To show this, we adopt a kinematic approach where the momentum/velocity equation is neglected and follow the Kazantsev prescription \citep{Kaz68,BS05}.

We shall include electron inertia subsequently and demonstrate that the Kazantsev approach \emph{will} introduce non-trivial contributions. 

\subsection{Kazanstev approach to Hall MHD}\label{SSecAMHD}

Let us begin by rewriting Ohm's law (\ref{OhmLaw}) as follows:
\begin{equation} \label{OhmLawAV}
\frac{\p {\bf b}}{\p t} =  \nabla \times \left[{\bf v} \times {\bf b} - d_i\left(\nabla \times {\bf b}\right) \times {\bf b}\right],
\end{equation}
where we have ignored the effects of dissipation (as done in the paper eventually), although its inclusion does not alter our result since the resistive term is linear in ${\bf b}$ and has no ${\bf v}$ dependence. We shall only run through the salient steps, as a more detailed derivation can be found in \citet{Kaz68}. The velocity field is taken to be delta correlated in time
\begin{equation}
\langle{v_i\left({\bf x},t\right) v_j\left({\bf x + r},t'\right)}\rangle = v_{ij}({\bf r}) \delta(t-t'),
\end{equation}
where the second order tensor $v_{ij}$ is given by
\begin{equation}
v_{ij} = 2V(r) \delta_{ij} + U(r) \left(\delta_{ij} - \frac{r_i r_j}{r^2}\right),
\end{equation}
and the assumption of incompressibility dictates that $U = r \p V/\p r$. In the Fourier space, we find that
\begin{equation}
\langle{v_i\left({\bf k},t\right) v_j\left({\bf k'},t'\right)}\rangle = \bar{V}(k) \delta(t-t')  \left(\delta_{ij} - \frac{k_i k_j}{k^2}\right) \delta({\bf k} + {\bf k'}),
\end{equation}
where $\bar{V}(r) = 3V(r) + U(r)$, and $\bar{V}(k)$ is its Fourier transform. The magnetic field can be taken to be a stochastic process and decomposed into its mean and fluctuating components. Since it is also divergence-free, and is assumed to possess similar statistics, most of the preceding equations are equally valid.

However, for the model considered herein, there is no mean field component, i.e. we have deliberately shut it off. Thus, the fluctuating part of the field can be represented by the total field ${\bf b}$ as the mean field component is absent.
In this scenario, the governing equation for the fluctuating part is the same as that of (\ref{OhmLawAV}) except for two extra terms given by
\begin{eqnarray} \label{OhmLawAVFP}
\frac{\p {\bf b}}{\p t} &=&  \nabla \times \left[{\bf v} \times {\bf b} - d_i\left(\nabla \times {\bf b}\right) \times {\bf b}\right] \nonumber \\
&& - \nabla \times \langle{{\bf v} \times {\bf b} - d_i\left(\nabla \times {\bf b}\right) \times {\bf b}}\rangle
\end{eqnarray}
However, the extra terms (on the second line of the RHS) in (\ref{OhmLawAVFP}) do not matter when computing $\langle{b_i b_j}\rangle$ as a result of the identity $\langle{\langle{(...)}\rangle b_j}\rangle = \langle{(...)}\rangle \langle{b_j}\rangle = 0$. Thus, we are free to drop these extra terms, which effectively transforms (\ref{OhmLawAVFP}) into (\ref{OhmLawAV}). The first term on the RHS of the latter expression is just the ideal MHD induction equation. There are several ways to treat this term, of which the use of the Furutsu-Novikov formula \citep{Furu63,Novi65} constitutes a particularly elegant approach.

Now, consider the second term on the RHS of (\ref{OhmLawAV}), which corresponds to the Hall term. Most standard treatments that rely upon the Kazantsev prescription assume that the fluctuating components obey Gaussian statistics \citep{Kaz68,BS05,IK13}. We see that the Hall term involves two factors of $b$; there also exist some gradients, or factors of $k$ when operating in the Fourier space. Thus, when we compute $\langle{b_i b_j}\rangle$ the Hall term involves \emph{three} factors of $b$. The Isserlis theorem \citep{Iss16,Iss18} in mathematical statistics dictates that all such terms, with an odd number of Gaussian variables, vanish upon averaging.\footnote{The Isserlis theorem is well-known in quantum field theory, under the guise of Wick's theorem \citep{PS95,Wein95}.} Consequently, this implies that the Hall term does not play any further role in the derivation of the evolution equation for $\langle{b_i b_j}\rangle$. Consequently, one obtains the ideal MHD Kazantsev result at the end, and the Hall term is rendered trivial.

At this stage, it may appear as though much of the algebra present in the paper might be rendered redundant as per the above reasoning. However, we emphasize that the assumptions of perfect homogeneity, isotropy, Gaussian statistics, etc. are highly idealized approximations, and are very likely to be invalid for realistic physical systems. Secondly, we note that the Kazantsev treatment is entirely kinematic, whereas a fully self-consistent model must take into account the coupled dynamical equations (\ref{MomEqn}) and (\ref{OhmLaw}). For these reasons, the Hall term is likely to play a non-trivial role, thereby meriting the analysis undertaken in this paper. Lastly, we observe that the existence of a finite mean field (not treated here) also modifies the Kazantsev treatment and introduces non-zero Hall corrections. 

\subsection{A brief comment on Extended MHD}
Hall MHD and ideal MHD rely on the assumption that the electrons are effectively massless. When the finite electron mass is included, the Ohm's law is rendered much more complicated. The correct expression for the extended Ohm's Law is known to be
\begin{eqnarray} \label{XMHDOhm}
\frac{\p {\bf b}^\star}{\p t} &=& \nabla \times \left[{\bf v} \times {\bf b}^\star - d_i\left(\nabla \times {\bf b}\right) \times {\bf b}^\star\right] \nonumber \\
&& + d_e^2 \nabla \times \left[\left(\nabla \times {\bf b}\right) \times \left(\nabla \times {\bf v}\right)\right],
\end{eqnarray}
where incompressibility has been assumed and the expression is normalized in Alfv\'en units \citep{LMM15}. Here, $d_e = c/\omega_{pe}$ is the electron skin depth and ${\bf b}^\star = {\bf b} + d_e^2 \nabla \times \left(\nabla \times {\bf b}\right)$. Note that $m_e \rightarrow 0$ implies that $d_e \rightarrow 0$ and ${\bf b}^\star \rightarrow {\bf b}$. It is apparent that ${\bf b}^\star$ is a \emph{linear} function of ${\bf b}$, and we can indeed use the shorthand notation ${\bf b}^\star = \mathcal{L} {\bf b}$, where $\mathcal{L} = 1 - d_e^2 \Delta$ with $\Delta$ denoting the Laplacian.

Let us suppose that we carry out the same steps outlined in Appendix \ref{SSecAMHD}. Owing to the absence of the mean-field terms, we obtain (\ref{XMHDOhm}) for the fluctuating part of ${\bf b}^\star$ with some additional terms that vanish upon employing the relation $\langle{\langle{(...)}\rangle b_j}\rangle = \langle{(...)}\rangle \langle{b_j}\rangle = 0$; note that this identity was also invoked earlier. At this stage, a clarification regarding ${\bf b}^\star$ is necessary. We work in Fourier space as this was the approach in Kazantsev's original paper, which we have also adopted. In this instance, note that the relation between ${\bf b}^\star$ and ${\bf b}$ is considerably simpler since ${\bf b}^\star_k = \left(1 - d_e^2 k^2\right){\bf b}_k$, where $k$ denotes the Fourier component. With the assumption of Gaussian statistics, and the linear relationship between ${\bf b}^\star$ and ${\bf b}$, we can multiply (\ref{XMHDOhm}) with ${\bf b}$ to compute $M_{ij}^\star := \langle{b_j \mathcal{L} b_i + b_i \mathcal{L} b_j}\rangle$. We shall not go into the mathematical subtleties here, but the existence of a well-defined inverse for $\mathcal{L}$ would, in principle, enable us to use $M_{ij}^\star$ to compute $\langle{b_i b_j}\rangle$, which is the desired object of interest.

The second term on the RHS of (\ref{XMHDOhm}) does \emph{not} contribute since there are three factors of ${\bf b}$ that arise in the Fourier representation, and the Isserlis theorem states that such terms vanish. This is exactly analogous to the discussion earlier, where the Hall term did not contribute to the Kazantsev prescription. The first term on the RHS of (\ref{XMHDOhm}) does contribute since there will be two factors of ${\bf b}$, and this term is akin to the ideal MHD term in the Kazantsev treatment. The last term on the RHS of (\ref{XMHDOhm}) is the most interesting, since it will also pick up two factors of ${\bf b}$ and thereby does not vanish upon averaging. Hence, in the case of extended MHD, we find that the Kazantsev treatment does lead to non-trivial corrections. Moreover, we find that these corrections are $\mathcal{O}\left(\lambda_e^2\right)$, with $\lambda_e = d_e/L$, suggesting that the additional terms are very small for most of the conventional astrophysical systems.

In this Appendix, we have thus shown that the assumption of a zero mean-field as well as the other standard assumptions of the Kazantsev dynamo lead to the absence of any contributions from the Hall term. However, the inclusion of electron inertia by means of extended MHD has been shown to yield additional terms that are likely to be very small, but finite.

\subsection{Comparison between the two approaches} \label{AppA3}
We shall briefly delineate the key differences between the approach adopted in the main body of the paper, and the Kazantsev prescription discussed in the Appendix. In particular, we shall restrict ourselves to the limitations of the latter, and indicate how these do not occur in the former.

The first aspect of the Kazantsev model is that it is altogether kinematic. In other words, the velocity field is fixed (with certain statistical properties) and the backreactions of the Lorentz force are not accounted for. On the other hand, our model takes into account the dynamical equations for the velocity and the magnetic field. Secondly, the Kazantsev model assumes that the velocity and magnetic fields obey the properties of statistical homogeneity, isotropy and Gaussianity. The latter of the trio is, in fact, responsible for eliminating the Hall effects altogether as shown in Appendix \ref{SSecAMHD}. The Kazantsev model may prove to be fairly accurate on small scales, as it's been demonstrated in the recent simulations by \citet{NB14} that MRI turbulence is isotropic (anisotropic) on smaller (larger) scales.

A careful study of Sec. \ref{SecClosure} reveals that our model has its own assumptions, but we do not invoke the above properties at any point in our derivation. In Sec. \ref{SecCSCase}, we do suppose that the stresses are homogeneous, but this is not a fundamental assumption, and was primarily called upon to simplify the model, and compare it with the previous results in the literature such as \citet{KY95,GIO03,PCP06,PCP08}.

\section{Alfv\'en waves in Hall MHD and implications for turbulence} \label{AppB}
It has been known since the 1940s \citep{Wal44} that the non-linear Alfv\'en wave is an exact solution of ideal MHD. It takes on the form ${\bf v} = \pm {\bf b}$, implying that the amplitudes of the velocity and magnetic field fluctuations are equal, whilst they are parallel or anti-parallel to each other. In the subsequent decades since their discovery, Alfv\'en waves have played a key role in understanding MHD turbulence \citep{Bisk03} and especially in the realm of wave turbulence \citep{Gal09}. In particular, `collisions' between Alfv\'en wave packets are responsible for setting up the turbulent cascade via their (nonlinear) interactions; for this reason, they have been referred to as the ``fundamental building block of plasma turbulence'' \citep{HowN13}. 

An immediate consequence of studying nonlinear Alfv\'en waves in ideal MHD is that the waves exhibit an equipartition of energy, as they obey the relation $v^2 = b^2$. If we envision turbulence as solely mediated by these waves, which possess equipartition of energy, it is `natural' to discuss equipartition in ideal MHD turbulence. Of course, this argument is primarily heuristic, and we defer the reader to rigorous studies of MHD turbulence \citep{Bisk03} for further details. There are two questions that can now be posed in the context of Hall MHD: (i) Can one derive non-linear Alfv\'en waves in Hall MHD? (ii) If such waves do exist, do they exhibit equipartition?

The answer to (i) is yes, and the relevant solutions were constructed in \citet{MK05} and further studied by \citet{AY16}. Thus, one may expect the nonlinear Alfv\'en waves of Hall MHD to play an analogous role to their ideal MHD counterparts, which leads us to the second question. Quite interestingly, it turns out that (ii) has a negative answer. It was shown in \citet{MK05} that
\begin{equation} \label{BVHMHD}
    {\bf b} = \alpha_\pm {\bf v},
\end{equation}
where $\alpha_\pm$ is given by 
\begin{equation} \label{alphaPMrel}
    \alpha_\pm = -\frac{k}{2} \pm \sqrt{\frac{k^2}{4} + 1},
\end{equation}
where the Hall parameter was set to unity for the sake of simplicity; for a non-unitary value, a similar expression can be easily recovered. From (\ref{BVHMHD}) and (\ref{alphaPMrel}), two immediate differences with respect to the ideal MHD case stand out. The first, of course, is the absence of equipartition for arbitrary values of $k$, since $v^2 \neq b^2$. Secondly, one finds that the ratio of the wave energies is a function of the wavelength, in sharp contrast with the ideal MHD case. In other words, the `Alfvenization' seen in ideal MHD is replaced by a different phenomenon, sometimes termed as `whistlerization', in Hall MHD. We reiterate that the latter effect leads intrinsically to a lack of equipartition, and the wave turbulence is mediated by whistler waves (which gave rise to this moniker). A detailed discussion of these aspects can be found in \citet{DDKD00,GB03,MK05,Gal06}.

As a result, one is led to conclude that the lack of equipartition is manifest in nonlinear Hall Alfv\'en waves. We have observed the same feature in our model, in this qualitative wave `turbulence' picture, and also in large scale fields, as shown by numerical studies \citep{MGM03b,MGM05}. Furthermore, there is experimental evidence of this phenomenon in the solar wind, where Hall MHD has been employed as the base model, and in the recent Hall MRI simulations by \citet{BLF16}. Thus, we may argue that this trait appears to be fairly universal, insofar as Hall MHD is concerned. Nevertheless, we also wish to add a cautionary word that this statement is, at the present moment, only a surmise.

We shall round off this Appendix by observing that this `Alfv\'enization' of turbulence also appears in dynamos. The standard $\alpha$-effect for ideal MHD was first calculated in \citet{PFL76} and it was shown that it has the form $\langle{{\bf v}\cdot \left(\nabla \times {\bf v}\right) - {\bf b}\cdot \left(\nabla \times {\bf b}\right) \rangle}$. It is easy to verify that the Alfv\'enic state ${\bf v} = \pm {\bf b}$ leads to the vanishing of the $\alpha$-effect. On the other hand, we expect to show, in a forthcoming publication, that this situation is radically altered when the Hall term is introduced. We intend to demonstrate that the Alfv\'enic state does not suppress the $\alpha$-effect in Hall dynamos.

\label{lastpage}

\end{document}